\documentclass[a4paper,11pt]{article}
\usepackage{pos}

\usepackage[dvipsnames]{xcolor}
\usepackage{multirow}
\usepackage{tabularx}
\usepackage{booktabs}

\newcommand{\OpenLoops}{{\rmfamily\scshape OpenLoops}}

\newcommand{\refeq}[1]{eq.~\eqref{#1}}

\newcommand{\reffi}[1]{Fig.~\ref{#1}}

\newcommand{\refta}[1]{Table~\ref{#1}}

\newcommand{\refse}[1]{Section~\ref{#1}}

\newcommand{\citere}[1]{Ref.~\cite{#1}}
\newcommand{\citeres}[1]{Refs.~\cite{#1}}

\newcommand{\ie}{i.e.\ }

\newcommand{\f}[2]{\frac{#1}{#2}}
\newcommand{\sss}[1]{\scriptscriptstyle#1}

\newcommand{\nosss}[1]{#1}

\newcommand{\bea}{\begin{eqnarray}}
\newcommand{\eea}{\end{eqnarray}}
\newcommand{\be}{\begin{equation}}
\newcommand{\ee}{\end{equation}}
\newcommand{\ba}{\begin{align}}
\newcommand{\ea}{\end{align}}
\newcommand{\beas}{\begin{eqnarray*}}
\newcommand{\eeas}{\end{eqnarray*}}
\newcommand{\bes}{\begin{equation*}}
\newcommand{\ees}{\end{equation*}}
\newcommand{\bas}{\begin{align*}}
\newcommand{\eas}{\end{align*}}

\newcommand{\eps}{{\varepsilon}}

\newcommand{\lb}{\left(}
\newcommand{\rb}{\right)}

\newcommand{\Dbar}[1]{{D}_{\nosss{#1}}}
\newcommand{\Db}[2]{{D}_{\nosss{#1}}^{(#2)}}

\newcommand{\momp}[1]{p_{#1}}
\newcommand{\mass}[1]{m_{\nosss{#1}}}

\newcommand{\momq}{\bar{q}}
\newcommand{\tilq}{\tilde{q}}

\newcommand{\calC}{\mathcal{C}}

\newcommand{\calN}{\mathcal{N}}
\newcommand{\barN}{\bar{\mathcal{N}}}

\newcommand{\barM}{\bar{\mathcal{M}}}
\newcommand{\topo}{\mathcal{\tau}}
\newcommand{\topoconf}{\vec{n}}

\newcommand{\topoconfdelta}{\Delta\vec{n}}

\newcommand{\nind}[1]{n_{\sss{#1}}}
\newcommand{\tilnind}[1]{m_{\sss{#1}}}

\newcommand{\calV}{\mathcal{V}}

\newcommand{\fullampbar}[2]{{\barM}_{{#1},{#2}}}
\newcommand{\colfac}[2]{{C}_{{#1},{#2}}}
\newcommand{\vertex}[1]{\calV_{#1}}
\newcommand{\ri}{\mathrm i}
\newcommand{\rd}{\mathrm d}

\newcommand{\nextmom}{N_p}

\definecolor{bluemar}{rgb}{0,0,.5}
\definecolor{redmar}{rgb}{.8,0,0}
\definecolor{greenmar}{rgb}{0,.5,0}

\graphicspath{{./pdfdiagrams/},{./figs/}}

\title{Two-loop tensor integral reduction for automated tools}
\author*{Fabian Lange}
\author{Max F.~Zoller}

\affiliation{Physik-Institut, Universität Zürich,\\
Winterthurerstrasse 190, 8057 Zürich, Switzerland}

\affiliation{PSI Center for Neutron and Muon Sciences,\\
5232 Villigen PSI, Switzerland}

\emailAdd{fabian.lange@physik.uzh.ch}
\emailAdd{max.zoller@physik.uzh.ch}

\abstract{In order to exploit the full potential of the LHC and future colliders, high-precision calculations of a very wide range of observables are crucial.
Automated tools for next-to-next-to-leading order calculations of perturbative scattering amplitudes are therefore a highly desirable goal.
So far, we have developed several major ingredients of such a tool in the \OpenLoops{} framework.

In our approach we split the calculation of scattering amplitudes into three components: loop momentum tensor integrals, the corresponding process-dependent tensor coefficients, and the interplay of $(D-4)$-dimensional parts of the integrand with divergences of the integrals.
In these proceedings, we present a new recursive algorithm to reduce arbitrary two-loop tensor integrals to scalar integrals, which has been implemented into an efficient numerical tool.
We also implemented a first version of the subsequent reduction to master integrals, which allows for a full validation of the algorithm.
We present a first successful validation computation and discuss its dependence on the precision of the externally computed master integrals.}

\FullConference{Loops and Legs in Quantum Field Theory (LL2026)\\
12-17, April, 2026\\
Bayreuth, Germany\\}

\begin{document}
\maketitle

\section{Introduction}

Scattering amplitudes computed in perturbation theory are an essential ingredient for Monte Carlo simulations of collider processes.
Tree and one-loop amplitudes have been available from fully automated numerical tools, such as \OpenLoops{}~\cite{Cascioli:2011va,Buccioni:2019sur}, for many years.
The precision requirements of the LHC and future colliders, however, demand next-to-next-to-leading order calculations, involving two-loop amplitudes, for a wide range of processes. This makes the development of similar algorithms and tools up to two loops crucial.
Our automation strategy in the \OpenLoops{} framework is to split the calculation into three major ingredients: process-dependent tensor coefficients, tensor integrals, and process-independent counterterms.
Recently, we developed a new and general algorithm for the numerical reduction of arbitrary two-loop tensor integrals, which was first presented in the proceedings~\cite{Lange:2026tpd}.
It is particularly well-suited to be interfaced into the \OpenLoops{} framework.\footnote{For other recent approaches to the problem of integral reduction for scattering amplitudes see \citeres{Chen:2019wyb,Peraro:2019cjj,Peraro:2020sfm,Anastasiou:2023koq,Goode:2024mci,Goode:2024cfy,Bevilacqua:2025xms}.}
We briefly review our algorithm in \refse{se:recursive-reduction}, before turning to our strategy to handle the resulting scalar integrals in \refse{se:scalar-integrals}.
We then validate our algorithm and implementation at the hand of a simple example in \refse{se:validation}.

\section{Recursive reduction of two-loop tensor integrals}
\label{se:recursive-reduction}

We compute amplitudes in the 't~Hooft--Veltman scheme~\cite{tHooft:1972tcz}, where external wave functions and momenta are four-dimensional, while loop momenta $\momq_i$, metric tensors $\bar{g}^{\bar\mu\bar\nu}$ and Dirac matrices $\bar\gamma^{\bar\mu}$ inside loops are defined in $D=4-2\eps $ dimensions in order to regularise divergences in loop integrals.
We denote these $D$-dimensional quantities with a bar, their projection to four dimensions without a bar, and the $(D-4)$-dimensional difference between those with a tilde, \ie
\bea
\momq_i = q_i + \tilde{q}_i , \quad
\bar{g}^{\bar\mu\bar\nu} = {g}^{\mu\nu} + \tilde{g}^{\tilde\mu\tilde\nu} , \quad \text{and} \quad
\bar{\gamma}^{\bar\mu} = {\gamma}^{\mu} + \tilde{\gamma}^{\tilde\mu} .
\eea

\begin{figure}[htbp]
  \centering
  \includegraphics[width=0.45\textwidth]{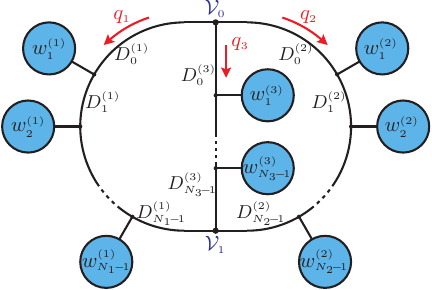}
  \caption{
    \label{fig:twoloopdia}
    General structure of an irreducible two-loop diagram.
    The blue blobs denote external subtrees.
    In general $\vertex{0}, \vertex{1}$ can be quartic vertices, in which case an external subtree is also attached there.
  }
\end{figure}

The amplitude of a generic two-loop diagram $\Gamma$ has the form
\be
\fullampbar{2}{\Gamma} =
\colfac{2}{\Gamma}
\int\!\rd\momq_1\int\!\rd\momq_2\,
\f{\barN(\momq_1,\momq_2)}
{\mathcal{D}(\momq_1,\momq_2)} \label{eq:intro_amp_two}\,,
\ee
where $\colfac{2}{\Gamma}$ is the colour factor of the diagram. The denominator
\be
\mathcal{D}(\momq_1,\momq_2)= \prod\limits_{i=1}^3 \prod\limits_{a=0}^{N_i-1} \Db{a}{i}(\momq_i) \quad \text{with} \quad \Db{a}{i}(\momq_i) = (\momq_i+p_{ia})^2-m_{ia}^2
\label{eq:den_prod_twoloop}
\ee
depends on the independent loop momenta $\momq_1,\momq_2$ or the linear combination \mbox{$\momq_3=-(\momq_1+\momq_2)$} as well as the propagator masses $m_{ia}$ and external momenta $p_{ia}$.
We choose the loop momenta in such a way that the number of propagators in \refeq{eq:den_prod_twoloop} is ordered as $N_1 \geq N_2 \geq N_3$.
For $N_3=0$ the two-loop diagram is reducible, \ie it factorises into two one-loop integrals.
An efficient algorithm for the construction of such amplitudes was presented in \citere{Pozzorini:2022ohr}.
The more challenging case of an irreducible diagram with $N_3\geq 1$ is depicted in \reffi{fig:twoloopdia}.
Its integrand factorises into three propagator chains $\calC_i$, each depending on a single loop momentum, connected by two vertices $\vertex{0},\vertex{1}$, which in general depend on both independent loop momenta $q_1,q_2$.

In our approach, we decompose the numerator of such a diagram into a four-dimensional part $\calN(q_1,q_2)=\barN(\momq_1,\momq_2)|_{D\to 4}$ and a remainder $\tilde{N}(\momq_1,\momq_2)$, which is of order $(D-4)$ and contributes only through its interplay with integral divergences.
We then apply a tensor decomposition to the four-dimensional numerator
\be
\calN(q_1,q_2) =
\sum\limits_{r=0}^{R_1}\sum\limits_{s=0}^{R_2(r)}\calN_{\mu_1\ldots\mu_r,\,\nu_1\ldots\nu_s}\,
q_1^{\mu_1}\ldots q_1^{\mu_r} q_2^{\nu_1}\ldots
q_2^{\nu_s}\, \label{eq:intro_tdec_two}
\ee
with upper bounds $R_i$ for the tensor ranks in the loop momenta $q_i$.
This decomposition leads to
\be
\fullampbar{2}{\Gamma} =
\colfac{2}{\Gamma} \lb \sum\limits_{r=0}^{R_1}\sum\limits_{s=0}^{R_2(r)}\calN_{\mu_1\ldots\mu_r,\,\nu_1\ldots\nu_s}\,
I_{\topo}^{\mu_1\ldots\mu_{r};\nu_1\ldots\nu_s}
+
\int\!\rd\momq_1\int\!\rd\momq_2\,
\f{\tilde{\calN}(\momq_1,\momq_2)}
{\mathcal{D}(\momq_1,\momq_2)}\rb
\label{eq:ampdia2L}
\ee
with two-loop tensor integrals of the form
\bea I_\tau^{\mu_1\ldots\mu_{r};\nu_1\ldots\nu_{s}}(\topoconf)
&=& \int\!\rd\momq_1\!\int\!\rd\momq_2
\f{q_1^{\mu_1}\cdots q_1^{\mu_{r}}\,q_2^{\nu_1}\cdots q_2^{\nu_{s}} \prod\limits_{i=1}^3 (\tilq_i^2)^{m_i}}{
 \prod\limits_{i=1}^3 \prod\limits_{a=0}^{N_i-1} \left[\Db{a}{i}(\momq_i)\right]^{\nind{ia}}} \quad\Big|_{q_3\to -(q_1+q_2)}{},
\label{eq:TIdia2l-ind}
\eea
where $\tau$ labels a given topology, and its configuration is defined as the set of indices
\be
\topoconf = (\vec{m},\topoconf_1,\topoconf_2,\topoconf_3) \label{eq:n_config_def_2l}
\ee
with $\vec{m} = (\tilnind{1},\tilnind{2},\tilnind{3})$ and $\topoconf_i = (\nind{i0},\ldots,\nind{i\,N_i-1})$.
For the integrals in \refeq{eq:ampdia2L} one has $\vec{m}=0$ and usually $\nind{ia}\in\{0,1,2\}$.

A highly efficient and numerically stable algorithm to recursively construct the tensor coefficients $\calN_{\mu_1\ldots\mu_r,\,\nu_1\ldots\nu_s}$ was presented in \citere{Pozzorini:2022ohr}.
Similarly as in the one-loop case~\cite{Ossola:2008xq,Draggiotis:2009yb,Garzelli:2009is,Pittau:2011qp}, the contributions associated with the interplay of $\tilde{\calN}$ with poles
can be reconstructed by means of two-loop rational counterterms~\cite{Pozzorini:2020hkx,Lang:2020nnl,Lang:2021hnw} together with the usual ultraviolet counterterms.
This approach is fully established for the interplay of $\tilde{\calN}$ with ultraviolet poles, while the rational terms of infrared origin are still under investigation~\cite{Zhang:2022rft}.

While the recursive construction of the tensor coefficients is performed purely numerically, our aim is to reduce the tensor integrals in a semi-numerical way, keeping the dependence on the loop momenta analytic until the computation of the master integrals in the last step, while computing reduction coefficients numerically.
In this spirit, we recently developed
a reduction formula for the generic tensor integral~\eqref{eq:TIdia2l-ind}, which was already presented in the proceedings~\cite{Lange:2026tpd}.
It exploits the exact identity
\bea q^\mu q^\nu
 &=&
 \sum\limits_{a=-2}^{3} \lb
  A^{\mu\nu}_{a} +B^{\mu\nu}_{a,\lambda} \,q^{\lambda} \rb \Dbar{a}(q)
\label{eq:qmuqnuredfinal}
\eea
for a four-dimensional loop-momentum tensor of rank two~\cite{delAguila:2004nf}, where the coefficients $A^{\mu\nu}_{a}$ and $B_{a,\lambda}^{\mu\nu}$ on the RHS are composed of three independent external momenta $p_{1},p_{2},p_{3}$, metric tensors, and the masses $\mass{a}$ of the propagators
\bea
 \Dbar{a}(\momq) = \begin{cases}
                        (\momq + \momp{a})^2-\mass{a}^2 & \text{ for } a \geq 0 , \\
                        1 & \text{ for } a = -1 ,\\
                        \tilq^2 & \text{ for } a = -2 .
                      \end{cases}
                      \label{eq:generalprop}
\eea
This allowed us to derive the reduction formula
\bea
I_\tau^{\mu_1\ldots\mu_{r};\nu_1\ldots\nu_{s}}(\topoconf)
&=&
\sum\limits_{\Omega\in\Gamma_r} \sum\limits_{\Omega'\in\Gamma_s}
\Bigg[
  C^{(0)\,\mu_1\ldots\mu_{r};\nu_1\ldots\nu_{s}}_{\Omega,\Omega'} I_{\topo}(\topoconf+\topoconfdelta_{\Omega,\Omega'}) \nonumber\\
  &&\hspace*{5em} + \sum\limits_{i=1}^{2}\sum\limits_{k=-1}^{\nextmom}
  C^{(i)\,\mu_1\ldots\mu_{r};\nu_1\ldots\nu_{s}}_{\Omega,\Omega',k} I_{\topo}\left(\topoconf+\topoconfdelta_{\Omega,\Omega'}+\topoconfdelta^{(i)}_{k}\right)\nonumber\\
  &&\hspace*{5em}
  + \sum\limits_{k=-2}^{3+4\nextmom+\nextmom^2}
  C^{(3)\,\mu_1\ldots\mu_{r};\nu_1\ldots\nu_{s}}_{\Omega,\Omega',k} I_{\topo}\left(\topoconf+\topoconfdelta_{\Omega,\Omega'}+\topoconfdelta^{(3)}_{k}\right)
\Bigg]
\label{eq:TIintegranl2l_fullred_toRank0}
\eea
with the reduction coefficients
\bea
  C^{(0)\,\mu_1\ldots\mu_{r};\nu_1\ldots\nu_{s}}_{\Omega,\Omega'} &=& A^{\mu_1 \cdots \mu_r}_{1,\Omega} A^{\nu_1 \cdots \nu_s}_{2,\Omega'} ,\nonumber\\
  C^{(1)\,\mu_1\ldots\mu_{r};\nu_1\ldots\nu_{s}}_{\Omega,\Omega',k} &=& A^{\nu_1 \cdots \nu_s}_{2,\Omega'} B^{\mu_1 \cdots \mu_r}_{1,\Omega, \lambda} C_{k}^{(1)\,\lambda} ,\nonumber\\
  C^{(2)\,\mu_1\ldots\mu_{r};\nu_1\ldots\nu_{s}}_{\Omega,\Omega',k} &=& A^{\mu_1 \cdots \mu_r}_{1,\Omega} B^{\nu_1 \cdots \nu_s}_{2,\Omega', \lambda'} C_{k}^{(2)\,\lambda'} ,\nonumber\\
  C^{(3)\,\mu_1\ldots\mu_{r};\nu_1\ldots\nu_{s}}_{\Omega,\Omega',k} &=& B^{\mu_1 \cdots \mu_r}_{1,\Omega, \lambda} B^{\nu_1 \cdots \nu_s}_{2,\Omega', \lambda'} C_{k}^{(3)\,\lambda\lambda'} .
\label{eq:TIintegranl2l_fullred_toRank0_coeff_ck}
\eea
The vectors $\topoconfdelta$ denote configuration shifts of the propagators, i.e.\ the appearance of propagator denominators in the numerator of the integrand.
The reduction coefficients $A^{\mu_1 \cdots \mu_r}_{i,\Omega}$ and $B^{\mu_1 \cdots \mu_r}_{i,\Omega, \lambda}$ are computed recursively from the rank-2 coefficients $A^{\mu\nu}_{a}$ and $B_{a,\lambda}^{\mu\nu}$ of \refeq{eq:qmuqnuredfinal}.
The first two sums run over the sets of all possible configuration shifts resulting from the recursive application of \refeq{eq:qmuqnuredfinal} to the rank-r and rank-s tensors in $q_1$ and $q_2$, respectively.
These sets are defined as
\bea
  \Gamma_r &=& \Big\{  \Omega = \{\omega_{-2}, \dots, \omega_{3}\} \;\;\Big|\;\;
  \omega_a \in \mathbb{N}_0,\;\;
  \sum\limits_a \omega_a = r-1
%   ,\;\;  \omega_{-2} \leq \f{r}{2}
  \Big\}. \label{eq:Gammar_def}
\eea
The coefficients $C_{k}^{(1)\,\lambda}$, $C_{k}^{(2)\,\lambda'}$, and $C_{k}^{(3)\,\lambda\lambda'}$ encode the remaining rank-1 and rank-1 $\times$ rank-1 reduction, not covered by this recursion.
They are obtained via Passarino-Veltman reduction~\cite{Passarino:1978jh}, for which $\nextmom$ denotes the number of external momenta.
We refer to our previous proceedings~\cite{Lange:2026tpd} and our forthcoming publication~\cite{Lange:2026XXX} for more details.

Following the \OpenLoops{} approach~\cite{Cascioli:2011va,Buccioni:2019sur}, we have implemented this algorithm in terms of a \texttt{Mathematica}~\cite{Mathematica} generator for the analytic steps and a \texttt{Fortran} code for an efficient numerical evaluation for each phase-space point.

\section{Handling scalar integrals}
\label{se:scalar-integrals}

After \refeq{eq:TIintegranl2l_fullred_toRank0}, we are left with the scalar integrals $I_{\topo}\left(\topoconf+\topoconfdelta_{\Omega,\Omega'}+\topoconfdelta^{(i)}_{k}\right)$.
In our implementation we follow the standard approach of an integration-by-parts reduction~\cite{Tkachov:1981wb,Chetyrkin:1981qh} with the Laporta algorithm~\cite{Laporta:2000dsw} and a subsequent evaluation of the master integrals.

Currently, we obtain the reduction tables analytically at the level of our \texttt{Mathematica} generator.
In a first step, we identify the master integrals and then reduce sample integrals with \texttt{Kira}~\cite{Maierhofer:2017gsa,Klappert:2020nbg,Lange:2025fba} with \texttt{FireFly}~\cite{Klappert:2019emp,Klappert:2020aqs} as backend.
This allows us to improve the master integral basis with \texttt{ImproveMasters.m}~\cite{Smirnov:2020quc}.
In particular, we search for a basis in which the space-time dependence factorises from the kinematic variables~\cite{Smirnov:2020quc,Usovitsch:2020jrk} and fewer spurious poles in $\epsilon$ appear.
Both requirements simplify the expressions in the reduction tables, and the latter in addition allows us to compute the master integrals to lower orders in $\epsilon$.
In a second step, we perform the reduction for the scalar integrals in \refeq{eq:TIintegranl2l_fullred_toRank0}, expand the resulting reduction tables with \texttt{Mathematica}, and translate them into \texttt{Fortran} code for the numerical evaluation.

Currently, the master integrals are evaluated with \textsc{pySecDec}~\cite{Borowka:2017idc,Borowka:2018goh,Heinrich:2021dbf,Heinrich:2023til}.
Their code is prepared by the generator, which takes into account the required depth for the $\epsilon$ expansion extracted from the reduction tables.
Then \textsc{pySecDec} is called for every phase-space point.

The emphasis of our present implementation is to verify the recursive reduction formula~\eqref{eq:TIintegranl2l_fullred_toRank0}.
Afterwards, we plan to explore possibilities to optimise the reduction and master integral pipeline, e.g.\ by shifting the integration-by-parts reduction to the numerical code, investigating further master integral tools, and weighting the requested precision of the master integrals by their prefactors.

\section{Validation}
\label{se:validation}

In our previous proceedings~\cite{Lange:2026tpd}, we already validated our recursive reduction down to rank one in each loop momentum at the integrand level.
More extensive validations of this step will be presented in our forthcoming publication~\cite{Lange:2026XXX}.

\begin{figure}[htbp]
  \centering
  \includegraphics[height=0.28\textwidth]{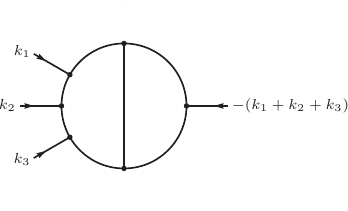}
  \caption{
    Two-loop pentagon-triangle topology used to validate our implementation.
  }
  \label{fig:pentagon-triangle}
\end{figure}

To validate our full algorithm and implementation, including the integral-level reduction for rank-1 and rank-1 $\times$ rank-1 integrals, we consider the $2\to2$ pentagon-triangle topology depicted in \reffi{fig:pentagon-triangle}.
We can express the required Mandelstam variables as
\begin{equation}
  s = (k_1 + k_2)^2 \quad \text{and} \quad t = (k_1 + k_3)^2
\end{equation}
in terms of the incoming external momenta $k_i$.
To regulate infrared divergences we set
\begin{equation}
  k_i^2 = m^2 .
\end{equation}
We then choose the topology of propagators of \refeq{eq:den_prod_twoloop} to be
\bea
\topo &=& \Big\{
\Db{0}{1},\Db{1}{1},\Db{2}{1},\Db{3}{1},\Db{0}{2},\Db{1}{2},\Db{2}{2},\Db{3}{2},\Db{0}{3}
\Big\}
\eea
with
\begin{alignat}{7}
  p_{01} &= 0,\quad
  &p_{11} &= k_1,\quad
  &p_{12} &= k_1 + k_2,\quad
  &p_{13} &= k_1 + k_2 + k_3,\nonumber \\
  p_{20} &= 0,\quad
  &p_{21} &= - k_1 - k_2 - k_3,\quad
  &p_{22} &= k_1,\quad
  &p_{23} &= k_1 + k_2, \\
  p_{30} &= 0 &&&&&& \nonumber
\end{alignat}
and all internal lines massless, i.e.\ $m_{ia} = 0$.
Hence, setting $s=1$, the process is described by the two kinematic variables $t$ and $m^2$.

As sample integral we choose the rank-4 top-level integral
\bea
I_\tau^{\mu_1\mu_{2}\mu_{3};\nu_1}(\vec{0},(1,1,1,1),(1,1,0,0),(1))
&=& \int\!\rd\momq_1\!\int\!\rd\momq_2
\f{q_1^{\mu_1} q_1^{\mu_{2}} q_1^{\mu_{3}} \, q_2^{\nu_1}}{
 \prod\limits_{i=1}^3 \prod\limits_{a=0}^{N_i-1} \left[\Db{a}{i}(\momq_i)\right]^{\nind{ia}}} ,
\eea
which is ultraviolet finite by power counting and infrared finite by the massive external legs.
Therefore, any $(D-4)$-dimensional $\tilq^2$ terms produced by our reduction do not contribute.
The rank of this integral is the minimal rank required to validate four parts of our algorithm:
the recursive reduction to rank one in each loop momentum, the Passarino-Veltman reduction to scalar integrals, the integration-by-parts reduction to master integrals, and the evaluation of the latter.

As reference value, we contract $I_\tau^{\mu_1\mu_{2}\mu_{3};\nu_1}(\vec{0},(1,1,1,1),(1,1,0,0),(1))$ with $p_{12,\mu_1} p_{12,\mu_2} p_{12,\mu_3} p_{12,\nu_1}$.
This allows us to rewrite
\begin{equation}
  p_{12,\mu} q_1^{\mu} = \frac{1}{2} \left[ \Db{2}{1} - \Db{0}{1} - s \right] \quad \text{and} \quad p_{12,\nu} q_2^{\nu} = p_{23,\nu} q_2^{\nu} = \frac{1}{2} \left[ \Db{3}{2} - \Db{0}{2} - s \right]
\end{equation}
in the integrand and obtain an analytic expression in terms of scalar integrals.
Those integrals are the same integrals that contribute to our reduction algorithm and are therefore included in the same reduction tables.
We evaluate the master integrals to $40$ digits with \texttt{AMFlow}~\cite{Liu:2022chg} with \texttt{Kira} as reduction backend, giving us an exact reference for a comparison with double precision results.

In \refta{tab:precision} we show the numerical precision at the physical phase-space point $t = -0.3$ and $m^2 = 0.1$ in dependence of the requested precision for the master integrals evaluated with \textsc{pySecDec}.
\begin{table}[htbp]
  \caption{
    Precision in dependence of the requested precision for the master integrals in \textsc{pySecDec} (MI prec).
    Correct digits are highlighted in green.
    The last line gives the reference value (Ref) derived with an analytic reduction and master integrals evaluated to $40$ digits with \texttt{AMFlow}.
  }
  \label{tab:precision}
  \begin{center}
    \begin{tabular}{c|c|c|c}
      \toprule
      MI prec & $\frac{1}{\epsilon^2}$ & $\frac{1}{\epsilon}$ & $\epsilon^0$ \\
      \midrule
      $10^{-4}$ & $-{\color{ForestGreen}0.00}36 - \ri\, {\color{ForestGreen}0.00}13$ & ${\color{ForestGreen}0}.13 + \ri\, {\color{ForestGreen}0}.28$ & ${\color{ForestGreen}3}5.5 + \ri\, {\color{ForestGreen}5}2.3$ \\
      $10^{-5}$ & $-{\color{ForestGreen}0.000}55 + \ri\, {\color{ForestGreen}0.000}33$ & ${\color{ForestGreen}0.0}34 - \ri\, {\color{ForestGreen}0.0}10$ & ${\color{ForestGreen}32}.99 + \ri\, {\color{ForestGreen}53}.58$ \\
      $10^{-6}$ & ${\color{ForestGreen}0.0000}39 + \ri\, {\color{ForestGreen}0.0000}82$ & ${\color{ForestGreen}0.000}30 + \ri\, {\color{ForestGreen}0.000}37$ & ${\color{ForestGreen}32.7}86 + \ri\, {\color{ForestGreen}53.47}91$ \\
      $10^{-7}$ & ${\color{ForestGreen}0.00000}70 + \ri\, {\color{ForestGreen}0.00000}83$ & ${\color{ForestGreen}0.0000}51 - \ri\, {\color{ForestGreen}0.000}12$ & ${\color{ForestGreen}32.76}85 + \ri\, {\color{ForestGreen}53.474}13$ \\
      $10^{-8}$ & ${\color{ForestGreen}0.0000000}24 - \ri\, {\color{ForestGreen}0.0000000}54$ & ${\color{ForestGreen}0.00000}19 + \ri\, {\color{ForestGreen}0.0000}10$ & ${\color{ForestGreen}32.7646}66 + \ri\, {\color{ForestGreen}53.4749}66$ \\
      \midrule
      Ref & $0$ & $0$ & $32.764695 + \ri\, 53.474944$ \\
      \bottomrule
    \end{tabular}
  \end{center}
\end{table}
The correct digits in green increase linearly with the requested number of digits, indicating that our algorithm is limited by the precision of the master integrals.
All reduction steps before as well as the linear combination cost about three digits for the finite term.

While the reduction steps of our algorithm and the evaluation of the reduction tables to master integrals only take a few milliseconds on a laptop, the evaluation of the master integrals with \textsc{pySecDec} takes between about one minute for a target precision of $10^{-4}$ to almost four hours for a target precision of $10^{-8}$.
For comparison, the numerical evaluation to $40$ digits with \texttt{AMFlow} took about $20$ minutes on the same laptop.
While this is sufficient for a validation of our algorithm and implementation, this problem has to be addressed for production code.

\section{Conclusions}

In these proceedings we summarized our new and general algorithm for the numerical reduction of arbitrary two-loop tensor integrals to scalar integrals first presented in the proceedings~\cite{Lange:2026tpd}.
It is based on a recursive reduction to rank one in each of the loop momenta, followed by a Passarino-Veltman reduction of the low-rank integrals to scalar integrals.
We discussed our strategy to handle the scalar integrals, namely an integration-by-parts reduction followed by a numerical evaluation of the master integrals with standard tools.
We validated our implementation at the hand of a simple example and showed that our code is currently limited by the master integrals, both in terms of runtime and precision.
On the other hand, the reduction steps of our algorithm and the reduction to master integrals are both fast and precise.
More details and further validations will be discussed in the forthcoming publication~\cite{Lange:2026XXX}.
Afterwards, the next step is to optimise our tool chain for the integration-by-parts reduction and the master integral evaluation to make our code ready for application in numerical tools, such as \OpenLoops{}.

\acknowledgments

This work was supported by the Swiss National Science Foundation (SNSF) under contract \href{https://data.snf.ch/grants/grant/211209}{TMSGI2\_211209}.

\bibliographystyle{JHEP}
\bibliography{bib}

\end{document}